\documentclass[aps,prl,twocolumn]{revtex4}
\usepackage{float}
\usepackage{makeidx}

\usepackage{graphicx}
\usepackage{amsmath}

\begin{document}

\title{Universal relations of strongly interacting Fermi gases with multiple scattering channels}

\author{Ran Qi}
\email{qiran@ruc.edu.cn}
\affiliation{Department of Physics, Renmin University of China, Beijing, 100872, P. R. China}

\begin{abstract}
Universal relations are important for understanding strongly interacting Fermi gases, the study of which have been mostly limited to cases with a single scattering channel. Here we discover a series universal relations for strongly interacting Fermi gases with multiple scattering channels. Unlike its counterpart across a single channel wide magnetic Feshbach resonance, a new kind of contact which we call the cross-channel contact naturally appears in this system, this contact is related to varies thermodynamic quantities as well as short range correlation functions.
\end{abstract}

\maketitle

In recent years, a number of new kind of experimental methods such as magnetic and optical
Feshbach resonances were invented to tune the inter-atomic interactions for alkali and alkaline-earth atoms \cite{Feshbach,Feshbach2}.
A vast number of efforts were devoted to investigate the strongly interacting state of matter in such systems \cite{manybody1,manybody2}.
Two most widely used theoretical models in those studies are the so called single channel model and two channel model in the spirit of effective field theory\cite{manybody1,twochannel}.

\begin{figure}[t]
\includegraphics[height=1.5 in, width=3.2 in]
{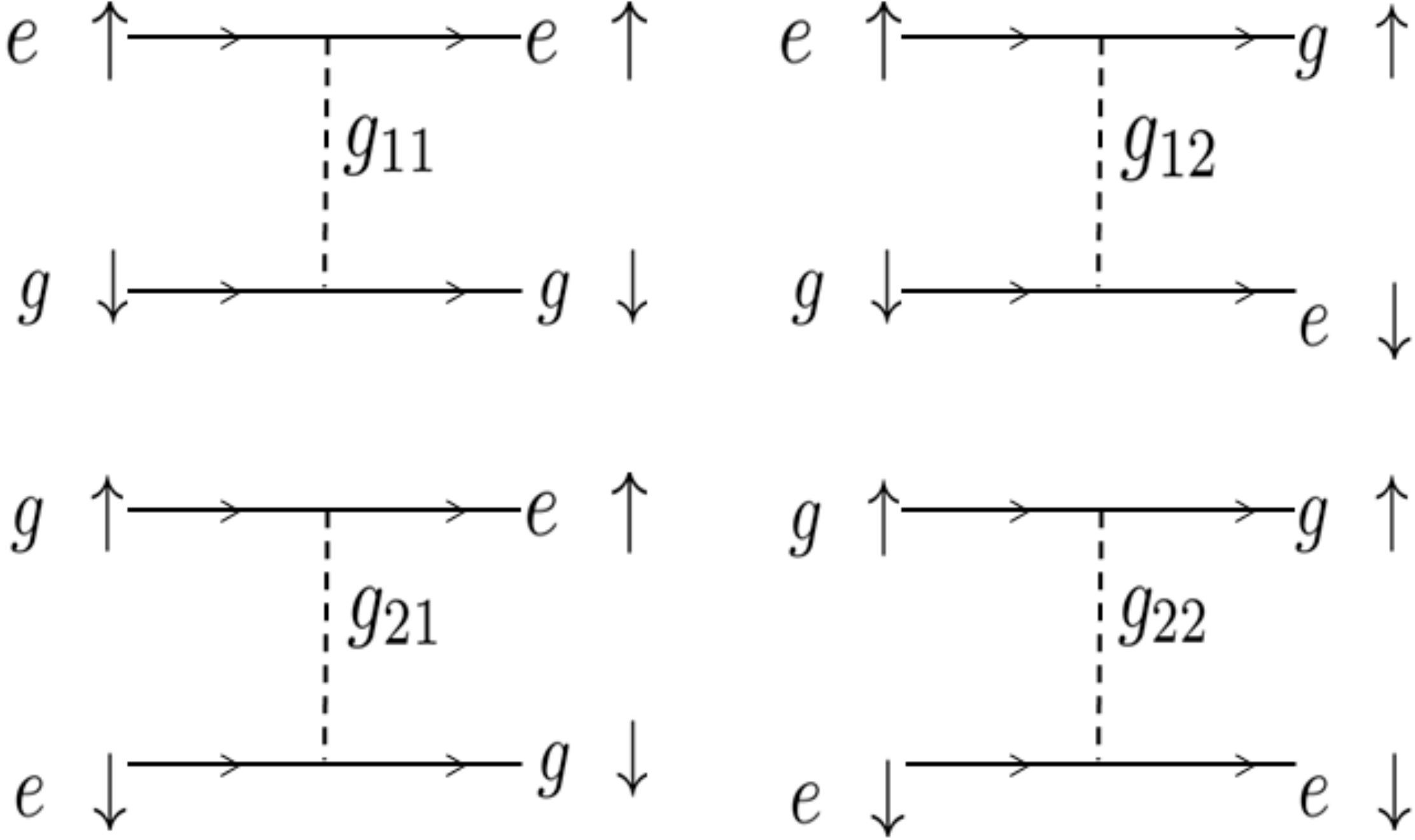}
\caption{Schematic scattering process of two alkaline-earth atoms across orbital Feshbach resonance. The coupling constants $g_{ij}$ are related to $g_{+}$ and $g_{-}$ in Eq. (\ref{interactionU}) as $g_{11}=g_{22}=(g_++g_-)/2$ and $g_{12}=g_{21}=(g_+-g_-)/2$.} \label{figOFR}
\end{figure}

In both cases, only a single scattering channel is introduced. The single channel model completely neglects the effects of closed channel molecules which is only valid for wide resonances with sufficiently small occupation on the closed channel. The two-channel model for narrow resonances, although included the coupling between open and closed channel, only treats the latter as a point like molecular state, and ignores all the scattering states in the closed channel. Such a treatment is only correct when the energy difference between the thresholds of the closed and the open channels are much larger than any relevant many-body energy scales (such as temperature and Fermi energy). This condition is full filled for most magnetic Feshbach resonances where the binding energy of closed channel molecules and the threshold detuning are usually  on the order of van de Waals energy. However, it breaks down in certain circumstances, such as the recently discovered orbital induced Feshbach resonances (OFR) in alkaline-earth atoms \cite{OFR1,OFR2,OFR3,OFR4}, as well as strongly interacting quantum gas with high spins \cite{highspin1,highspin2,highspin3,highspin4}. For example, across the OFR, the system has a very small detuning energy around the resonance due to the extremely small nuclear spin Zeeman splitting energy. This detuning energy is comparable to the Fermi energy of the system under experimentally reasonable densities, for which the scattering effects in closed channel is nonnegligible. Further more, the corresponding closed channel molecule is very shallow whose size is as large as about $2000a_0$ which is comparable to average inter-particle distance under typical atomic densities and can not be treated as point like particles. As a result, a more complete description which properly takes into account the scattering states of both open and closed channels is required.

On the other hand, the existence of strong correlation effects and the lack of small parameters for perturbative calculations makes rigorous results very precious for such strongly interacting systems. In recent years, a series of exact universal relations was established which applies to all single channel many-body systems \cite{contact,contact2} and was later generalized to the two-channel model \cite{contact3}. As mentioned above, such models are only suitable for systems with sufficiently large detuning energy where only a single {\it scattering} channel is effectively open. A natural question would be that how to generalize such relations to a system with multiple {\it scattering} channels such as alkaline-earth atoms across OFR or strongly interacting Fermi gas with high spins.

In this Letter, we generalize such exact relations to a system with multiple {\it scattering} channels. By implementing a ``multi-band model" we properly included the scattering effects for both open and closed channels \cite{OFR1,twoband}. We established a series of universal relations with the help of operator product expansion (OPE) for quantum fields. In the earlier works \cite{contact,contact2,contact3,contact4,contact5,contact6,contact7}, a key quantity called the ``Contact" relates the short range correlation of the many-body wave function to the macroscopic thermodynamic quantities. One of the most important properties of the Contact is that it is always related to the $1/k^4$ behavior in the large momentum tail of the momentum distribution. For a system with multiple {\it scattering} channels considered in this Letter, we found that instead of a single Contact one needs to introduce a ``Contact matrix" in order to establish the universal relations. The diagonal components of this Contact matrix, which we call the intra-channel Contacts, are indeed related to the $1/k^4$ tail of momentum distributions. There is, however, a new kind of Contact corresponding to the off-diagonal components which does not relate to any large momentum tails but still appears naturally in a sequence of universal relations. We call it ``cross-channel Contact" since it is defined through the matrix element of pairing operators across different scattering channels. In the following, we first introduce our model and the definition of our Contacts and then establish the universal relations.

The scattering between two alkaline-earth atoms is shown schematically in Fig. \ref{figOFR}, where $|g\rangle$ and $|e\rangle$ denote the electronic ground state $^1S_0$ and excited state $^3P_0$ while $|\uparrow\rangle$ and $|\downarrow\rangle$ corresponds to two nuclear spin states $m_I$ and $m_{I'}$ \cite{OFR1}.
The many-body system can be well captured by the following Hamiltonian
\begin{eqnarray}
\hat{H}&=&\hat{H}_{0o}+\hat{H}_{0c}+\hat{U}+\hat{V}_{ext},\label{hamiltonian}\\
\hat{H}_{0o}&=&\sum_{\mathbf{k}}\epsilon_{\mathbf{k}}(\alpha_{1\mathbf{k}}^{\dag}\alpha_{1\mathbf{k}}+\beta_{1\mathbf{k}}^{\dag}\beta_{1\mathbf{k}}),\\
\hat{H}_{0c}&=&\sum_{\mathbf{k}}\left(\epsilon_{\mathbf{k}}+\frac{\delta}{2}\right)(\alpha_{2\mathbf{k}}^{\dag}\alpha_{2\mathbf{k}}+\beta_{2\mathbf{k}}^{\dag}\beta_{2\mathbf{k}}),
\end{eqnarray}
where $\alpha_{1},\beta_{1},\alpha_{2},\beta_{2}$ are the annihilation operator of $|e\uparrow\rangle,|g\downarrow\rangle,|g\uparrow\rangle,|e\downarrow\rangle$ states respectively and $\epsilon_{\mathbf{k}}=\hbar^2\mathbf{k}^2/(2m)$,
\begin{eqnarray}
\hat{U}&=&\int d^3\mathbf{r}\left[\frac{g_+}{2}\Pi_+^{\dag}(\mathbf{r})\Pi_+(\mathbf{r})+\frac{g_-}{2}\Pi_-^{\dag}(\mathbf{r})\Pi_-(\mathbf{r})\right]\label{interactionU}
\end{eqnarray}
represents the inter-particle interaction, and $\hat{V}_{ext}$ is some arbitrary external potential. To simplify the notation, we have defined two sets of pair operators: $\Pi_{\pm}(\mathbf{r})=\Pi_1(\mathbf{r})\mp\Pi_2(\mathbf{r})$ and $\Pi_i(\mathbf{r})=\alpha_i(\mathbf{r})\beta_i(\mathbf{r})$ where $i=1,2$ refers to open and closed scattering channels. We will also take natural units and set $m=\hbar=1$ in the following. The bare coupling constants $g_+$ and $g_-$ are related to two physical scattering lengths $a_+$ and $a_-$ with the following renormalization relations \cite{OFR1}
\begin{eqnarray}
\frac{1}{g_{\pm}}=\frac{1}{4\pi a_{\pm}}-\frac{\Lambda}{2\pi^2},\label{renormalization}
\end{eqnarray}
where $\Lambda$ is the momentum cut off. As illustrated in \cite{OFR1}, this Hamiltonian gives a two-body s-wave scattering length:
\begin{eqnarray}
a_s=\frac{a_+ + a_--2\sqrt{\delta}a_+ a_-}{2-\sqrt{\delta}(a_+ + a_-)}.\label{as}
\end{eqnarray}
The scattering resonance corresponds to $\delta=4/(a_+ + a_-)^2$ where $a_s$ diverges and the system enters a strongly interacting regime.

{\it Intra- and inter- channel Contacts.} Here we first define a $2\times 2$ Contact matrix as
\begin{eqnarray}
C_{ij}\equiv \left(\frac{\Lambda}{2\pi} \right)^2\int d^3\mathbf{R}\langle\Pi_i^{\dag}(\mathbf{R})\Pi_j(\mathbf{R})\rangle. \label{matrix}
\end{eqnarray}
With the help of OPE for the operator $\alpha_i^{\dag}(\mathbf{R+r/2})\alpha_i(\mathbf{R-r/2})$ and $\beta_i^{\dag}(\mathbf{R+r/2})\beta_i(\mathbf{R-r/2})$, we have the following asymptotic small $r$ expansion:
\begin{eqnarray}
&&\alpha_n^{\dag}(\mathbf{R+r/2})\alpha_n(\mathbf{R-r/2})=\nonumber\\
&&\alpha_n^{\dag}(\mathbf{R})\alpha_n(\mathbf{R})+\sum_{k,\ell} f^n_{k\ell}(r) \Pi_k^{\dag}(\mathbf{R})\Pi_{\ell}(\mathbf{R})+\cdots,\label{nkope}
\end{eqnarray}
where $f^n_{k\ell}(r)$ is a set of universal functions called the Wilson coefficients ($n,k,\ell=1,2$) and we only included the terms with leading order non-analytic behaviors. Since this is an operator equation, it should stands for the matrix element between two arbitrary states. Following a similar procedure in \cite{contact2}, we consider its matrix element between one two-body scattering state with both particles in channel $i$ (denote as $|p,i\rangle$) and another state with both particles in channel $j$ (denote as $|p,j\rangle$), where $p$ is the relative momentum. On one hand, the matrix elements of the operator $\Pi_k^{\dag}(\mathbf{R})\Pi_{\ell}(\mathbf{R})$ is given as
\begin{eqnarray}
\langle p,j|\Pi_k^{\dag}(\mathbf{R})\Pi_{\ell}(\mathbf{R})|p,i\rangle=\left(\frac{\Lambda}{2\pi^2}\right)^2 T_{ik}(p)T_{\ell j}(p).\label{rhs}
\end{eqnarray}
On the another hand, the matrix element of r.h.s in Eq. (\ref{nkope}) is given as
\begin{eqnarray}
\langle p,j|\alpha_n^{\dag}(\mathbf{R+r/2})\alpha_n(\mathbf{R-r/2})|p,i\rangle=i\frac{e^{ipr}}{8\pi}T_{in}(p)T_{n j}(p).\nonumber\\
~\label{lhs}
\end{eqnarray}
Substituting Eq. (\ref{rhs}), (\ref{lhs}) into (\ref{nkope}) and comparing the coefficients up to order $r$, one immediately finds $f^n_{k\ell}(r)=r\delta_{kn}\delta_{\ell n}$. After taking the fourier transform for Eq. (\ref{nkope}) one finally obtains
\begin{eqnarray}
n_{\alpha_1}(\mathbf{k}),n_{\beta_1}(\mathbf{k})\rightarrow\frac{C_{11}}{k^4},~\text{and}~
n_{\alpha_2}(\mathbf{k}),n_{\beta_2}(\mathbf{k})\rightarrow\frac{C_{22}}{k^4},\label{nk}
\end{eqnarray}
where the two intra-channel Contacts $C_{11}$ and $C_{22}$ are the diagonal components of an $2\times2$ Contact matrix defined in Eq. (\ref{matrix}). And we have shown that they relate to the $1/k^4$ tail of momentum distributions in both the open and the closed channels, respectively. For similar reason as in the single channel model, these two intra-channel Contact also relate to the high-frequency tail of the radio-frequency transition rate \cite{contact2}. Suppose a rf signal with
frequency $\omega$ is applied to transfer atoms from the $\alpha_i$ or $\beta_i$ to another empty state, then the transition rate $\Gamma_i(\omega)$ has the following asymptotic behavior in the limit $\omega\rightarrow\infty$:
\begin{eqnarray}
\Gamma_i(\omega)\rightarrow\frac{\Omega^2}{4\pi\omega^{3/2}}C_{ii},
\end{eqnarray}
where $i=1,2$ and $\Omega$ is the Rabi frequency determined by the strength of the rf signal.

However, as it will be clear later, the intra-channel Contacts $C_{ii}$ alone are not sufficient to derive the following universal relations. One also need the off-diagonal component $C_{cr}=C_{12}=C_{21}^*$
which we call ``cross-channel Contact". It is easy to show that one can always make $C_{12}$ real with a proper gauge choice and thus we will assume $C_{cr}=C_{cr}^*$
in the following part of this paper. Although this $C_{cr}$ does not relate to the $1/k^4$ tails in momentum distributions, it still plays an important role in the following exact relations including the energy relation, the adiabatic relations, the pressure relations, and the virial theorem.


{\it Energy relation.} The energy relation can be derived simply by using the renormalization relation in Eq. (\ref{renormalization}) to express the Hamiltonian
 in Eq. (\ref{hamiltonian}) as the sum of three terms whose matrix elements are ultraviolet finite. We then finally arrive at
\begin{eqnarray}
E\!=\!\sum_{\mathbf{k}}\!\epsilon_{\mathbf{k}}\!\left[n_{\mathbf{k}}\!-\!2\frac{C_{tot}}{k^4}\right]\!+\!\frac{N_c\delta}{2}\!+\!\bar{I}\!+\!\langle\hat{V}_{ext}\rangle, \label{energetic}
\end{eqnarray}
where $n_{\mathbf{k}}=\sum_i[n_{\alpha_i}(\mathbf{k})+n_{\beta_i}(\mathbf{k})]$ is the momentum distribution of total particle number, $C_{tot}=C_{11}+C_{22}$, $N_c=\sum_{\mathbf{k}}[n_{\alpha_2}(\mathbf{k})+n_{\beta_2}(\mathbf{k})]$ is the total occupation number in closed channel. The quantity $\bar{I}$ is related to both intra-channel and cross-channel Contacts:
\begin{eqnarray}
\bar{I}&\!=\!&\frac{1}{8\pi}\left[\!\left(\frac{1}{a_+}\!+\!\frac{1}{a_-}\! \right)C_{tot}\!+\!2\left(\frac{1}{a_-}\!-\!\frac{1}{a_+}\! \right)C_{cr}\right].
\end{eqnarray}

{\it Adiabatic relations.} The adiabatic relations can be derived by using the Feynman-Hellman theorem:
\begin{eqnarray}
dE/d\lambda=\langle \partial\hat{H}/\partial\lambda\rangle.
\end{eqnarray}
By choosing $\lambda$ as either one of the three two-body parameters $a_+,a_-$ and $\delta$, and noting that $\hat{H}$ only depends on $a_{\pm}$ through $g_{\pm}$ in Eq. (\ref{renormalization}), we finally obtain
\begin{eqnarray}
\left(\frac{\partial}{\partial a_+^{-1}}+\frac{\partial}{\partial a_-^{-1}}\right)E&=&-\frac{C_{tot}}{4\pi}\label{adiabatic1},\\
\left(\frac{\partial}{\partial a_-^{-1}}-\frac{\partial}{\partial a_+^{-1}}\right)E&=&-\frac{C_{cr}}{2\pi}\label{adiabatic2},\\
\frac{\partial}{\partial \delta}E&=&\frac{N_c}{2}.
\end{eqnarray}
At finite temperature, all the derivatives in the above relations must be taken at fixed entropy.

In a realistic OFR, it is usually very difficult to tune the value of $a_+$ and $a_-$ and direct application of the adiabatic relations (\ref{adiabatic1}) and (\ref{adiabatic2}) could be difficult. However, they are still useful in estimating the two-body loss rate \cite{contact2}. For example, if the fermions have a small inelastic two-body scattering amplitude, then the optical theorem implies that the two scattering lengths $a_{\pm}$ will have negative imaginary parts. The leading many-body effects of such an inelastic scattering process is to induce a small decay rate, which can now be estimated with the above adiabatic relations. Up to leading order in $\text{Im}a_{\pm}$, the decay rate $\Gamma$ is given as
\begin{eqnarray}
\Gamma\!\simeq\!\frac{1}{4\pi}\!\left[\!\frac{C_{tot}}{2}\!\left(\!\frac{\text{Im}a_-}{|a_-|^2}\!+\!\frac{\text{Im}a_+}{|a_+|^2}\!\right)
\!+\!C_{cr}\left(\!\frac{\text{Im}a_-}{|a_-|^2}\!-\!\frac{\text{Im}a_+}{|a_+|^2}\!\right)\!\right]\!
\end{eqnarray}
which is valid when $\text{Im}a_{\pm}\ll|a_{\pm}|$.

{\it Pressure relation.} For a uniform system with $\hat{V}_{ext}=0$, we can write the total energy at finite temperature in a scaling form
\begin{eqnarray}
E=V^{-\frac{2}{3}}F\left(\frac{a_+}{V^{\frac{1}{3}}},\frac{a_-}{V^{\frac{1}{3}}},V^{\frac{2}{3}}\delta,\frac{S}{k_B},N\right),
\end{eqnarray}
where $F$ is a dimensionless function, $N,V$ is the total particle number and volume, and $S$ is the entropy. By taking the derivative of $V$ together with the thermodynamic relation $P=-\partial E/\partial V|_S$, we
 obtain the following pressure relation
\begin{eqnarray}
PV=\frac{2}{3}E+\frac{1}{3}\bar{I}-\frac{1}{3}N_c\delta,
\end{eqnarray}
where $P$ is the pressure.

{\it Virial theorem.} For a trapped system with $\hat{V}_{ext}$ being an isotropic harmonic potential, we again write the total energy in the following scaling form
\begin{eqnarray}
E=\omega G\left(a_+\sqrt{\omega},a_-\sqrt{\omega},\frac{\delta}{\omega},\frac{S}{k_B},N\right),
\end{eqnarray}
where $G$ is a dimensionless function and $\omega$ is the trapping frequency. By taking the derivative of $\omega$ and using the relation $\partial E/\partial\omega=2\langle\hat{V}_{ext}\rangle$ from the Feynman-Hellman theorem, we finally arrive at the Virial theorem :
\begin{eqnarray}
E=2\langle\hat{V}_{ext}\rangle-\frac{1}{2}\bar{I}+\frac{1}{2}N_c\delta.
\end{eqnarray}

Above we have summarized our main results of varies universal relations. One can clearly see that both the intra channel and the crossed channel Contacts plays an important role in these relations, although the later one is not related to any large momentum tails. In the following, we analyze the behavior of the ground state Contacts in both BCS and BEC limit where simple analytical results can be obtained.

\begin{figure}[t]
\includegraphics[bb=47bp 279bp 542bp 504bp, height=1.6 in, width=3.4 in]
{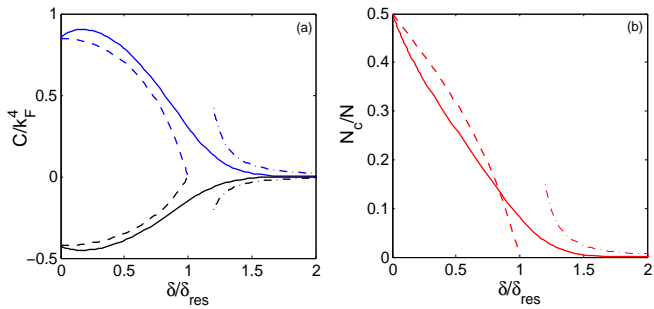}
\caption{(a) Intra-channel and cross-channel Contact $C_{tot}$ (blue lines) and $C_{cr}$ (black lines). (b) Closed channel fraction $N_c/N$. The solid, dashed and dotted line refers to meanfield results, BEC limit, and BCS limit asymptotic behaviors respectively. The parameters are chosen as $a_-=0.09a_{+},~k_Fa_+=0.5$. }\label{contact}
\end{figure}

{\it BCS limit.} In the weak coupling BCS limit, the leading order contribution for the ground state energy comes from the Hartree-Fock term which gives
\begin{eqnarray}
\frac{E}{V}=4\pi a_s n_{\alpha_1}n_{\beta_1},
\end{eqnarray}
where $a_s$ is the s-wave scattering length defined in (\ref{as}) and $n_{\alpha_1},~n_{\beta_1}$ is the number density of $\alpha_1$ and $\beta_1$ particles. Since all the dependence on $a_{\pm}$ and $\delta$ only comes from $a_s$, one can easily obtain the Contacts and closed channel occupation number by applying the adiabatic relations:
\begin{eqnarray}
\frac{C_{tot}}{V}&=&16\pi^2n_{\alpha_1}n_{\beta_1}(a_s^2+a_{12}^2),\\
\frac{C_{cr}}{V}&=&16\pi^2n_{\alpha_1}n_{\beta_1}a_sa_{12},\\
\frac{N_c}{V}&=&4\pi n_{\alpha_1}n_{\beta_1}\frac{a_{12}^2}{\sqrt{\delta}},
\end{eqnarray}
where $a_{12}$ is the inter channel scattering length given as
\begin{eqnarray}
a_{12}=\frac{a_--a_+}{2-\sqrt{\delta}(a_+ + a_-)}\label{a12}.
\end{eqnarray}

{\it BEC limit.} On the other hand in the BEC limit, the ground state is a nearly noninteracting Bose gas of Feshbach molecules leading to:
\begin{eqnarray}
\frac{E}{V}\simeq-nE_b,
\end{eqnarray}
where $n=n_{\alpha_1}+n_{\alpha_2}=n_{\beta_1}+n_{\beta_2}$ and we only consider the balanced case. $E_b$ is the absolute value of two body binding energy satisfying
\begin{eqnarray}
&&\frac{1}{a_+a_-}+\sqrt{E_b(E_b+\delta)}+\nonumber\\
&&-\frac{1}{2}\left(\frac{1}{a_+}+\frac{1}{a_-} \right)\left(\sqrt{E_b}+\sqrt{E_b+\delta}\right)=0.
\end{eqnarray}
Now the dependence on $a_{\pm}$ and $\delta$ only comes from $E_b$ and we find
\begin{eqnarray}
\frac{C_{tot}}{V}&=&\frac{8\pi n}{\gamma}\left(\sqrt{E_b}+\sqrt{E_b+\delta}-\frac{1}{a_+}-\frac{1}{a_-}\right),\\
\frac{C_{cr}}{V}&=&\frac{4\pi n}{\gamma}\left(\frac{1}{a_-}-\frac{1}{a_+}\right),\\
\frac{N_c}{N}&=&\frac{1}{\gamma}\left[\sqrt{\frac{E_b}{E_b+\delta}}-\frac{\frac{1}{a_+}+\frac{1}{a_-}}{2\sqrt{E_b+\delta}}\right],
\end{eqnarray}
where the dimensionless $\gamma$ is defined as
\begin{eqnarray}
\gamma=\frac{2E_b+\delta}{\sqrt{E_b(E_b+\delta)}}\!-\!\frac{1}{2}\left(\!\frac{1}{a_+}\!+\!\frac{1}{a_-} \!\right)\left(\!\frac{1}{\sqrt{E_b}}\!+\!\frac{1}{\sqrt{E_b+\delta}}\!\right).\nonumber
\end{eqnarray}

Away from both limits, no exact results are known and we estimate the contact by performing a meanfield calculation based on the Hamiltonian (\ref{hamiltonian}) \cite{OFR1}. The results are shown in Fig. \ref{contact} where the dashed lines refers to analytic results in two limits. The atomic density and value of $a_{\pm}$ are taken from realistic values of current experiments \cite{OFR2,OFR3}. One can see that in the near resonance region as well as the BEC limit, the magnitude of cross-channel Contact is always comparable to the intra-channel Contact and thus can not be neglected with reasonable experimental setups. We have also shown that the fraction of closed channel occupation $N_c/N$ is of order 1 across the resonance which further implies that the system cannot be described by a single channel model and our generalized form of universal relations are necessary for such system.

{\it Outlook.} Although we only took the strongly interacting Fermi gas across OFR as an example to illustrate our results. The approach we used and the qualitative structures of the universal relations can be easily generalized to any system with arbitrary number of scattering channels. In general, for a system with $N$ different channels there should be an $N\times N$ Contact matrix associated with the corresponding universal relations while only the $N$ diagonal components are related to the $1/k^4$ large momentum tails. The off diagonal ones although does not relate to the momentum tails will appear in different universal relation which can be verified experimentally. These results are very different from the relations with the currently widely used single channel and two-channel models. Our results for the alkaline-Earth atoms could be verified with the ongoing $^{173}$Yb experiments with the help of OFR \cite{OFR2,OFR3}.

{\it Acknowledgements.} We thank Hui Zhai, Peng Zhang, Shina Tan, Zhenhua Yu, Xiaji Liu and Hui Hu for useful discussions. This work is supported by the Fundamental Research Funds for the Central Universities, and the Research Funds of Renmin University of China under Grant No. 15XNLF18 and No. 16XNLQ03.

\end{document}